# Measuring the ϒ Nuclear Modification Factor at STAR

**Rosi Reed for the STAR Collaboration**
University of California, Davis
E-mail: rjreed@ucdavis.edu

**Abstract.** Suppression of quarkonia in heavy ion collisions with respect to proton-proton collisions due to the Debye screening of the potential between the heavy quarks was hypothesized to be a signature of the Quark Gluon Plasma (QGP) [1]. However, other effects besides Debye screening, such as the statistical recombination of heavy flavor $Q\bar{Q}$ pairs, or co-mover absorption can also affect quarkonia production in heavy ion collisions. Quantifying the suppression of an entire family of quarkonium mesons can give us a model dependent constraint on the temperature. The suppression of ϒ can be quantified by calculating the Nuclear Modification factor, $R_{AA}$, which is the ratio of the production in Au+Au collisions to the production in p+p scaled by the number of binary collisions. We present our results for mid-rapidity ϒ(1S+2S+3S) production in p+p and Au+Au collisions at $\sqrt{s_{NN}}$ = 200 GeV. The centrality dependence of $R_{AA}$ will be shown for the combined ϒ(1S+2S+3S) yield.

## 1. Introduction

Suppression of quarkonia due to the Debye screening of the potential between the two heavy quarks was thought to be a distinct signature of QGP formation [1]. Suppression was observed in heavy ion collisions at both SPS and RHIC [2]. The magnitude of suppression in both systems was similar despite the different energy densities, indicating that other physics processes besides Debye screening needed to be accounted for. An important measurement that can be made by studying the ϒ 1S, 2S, and 3S states is to constrain the QGP temperature in a model dependent way. At 200 GeV, calculations constrained by lattice data indicate that the ϒ(3S) should be completely dissociated, while the ϒ(2S) state may dissociate and the ϒ(1S) state should survive [3,4]. In order to measure the medium modification of the ϒ(1S, 2S, 3S) states a baseline p+p measurement of the cross-section was required and determined to be 114±38(stat)+23/-24 pb for $|y_ϒ|<0.5$ [4].

## 2. Determining ϒ yield in Au+Au Collisions

The data in these proceedings is from 50 Million events that satisfied the high tower trigger taken during the 2010 RHIC run. This is equivalent to $4.62 \times 10^9$ minimum bias triggers. The high tower trigger requires a single tower above 4.2 GeV in STAR's Barrel Electromagnetic Calorimeter (BEMC). Tracks in the Time Projection Chamber (TPC) were selected with an ionization energy loss that matched the expected electron ionization energy loss. Further particle identification was applied to each track by extrapolating the tracks to the BEMC and requiring that the E/p is close to 1 as expected for electrons. E is the energy left in a three tower cluster in the BEMC, and p is the momentum of the track as measured in the TPC. ϒ candidates were formed from track pairs, with the requirement that at least one of the tracks matches to a tower that could have fired the trigger. The like-sign and unlike-sign invariant mass spectra between 7 and 12 GeV/c$^2$ for 0-60% centrality with |y|<0.5 are shown in figure 1.

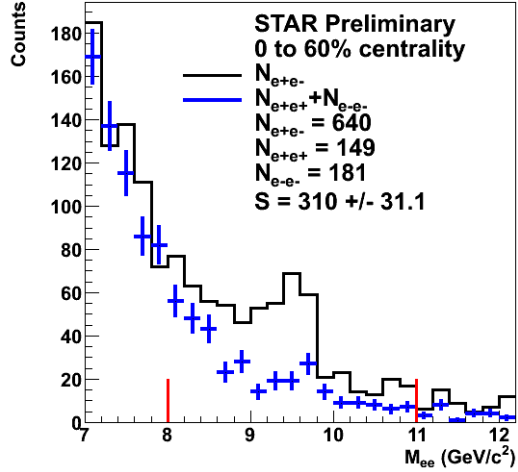

**Figure 1.** The like-sign invariant mass spectrum of dielectron pairs is shown in black (N$_{+-}$) and the unlike-sign spectrum is shown in blue (N$_{--}$ + N$_{++}$). The significance of the unlike-sign spectrum over the like-sign background is 10σ in the region between 8 and 11 GeV/c$^2$, as shown by the red lines.

Extracting the ϒ yield requires knowledge of the $b\bar{b}$ and Drell-Yan→e$^+$e$^-$ background. There are large theoretical uncertainties in these yields at M$_{ee}$ = 10 GeV/c$^2$ in p+p collisions. The amount that the $b\bar{b}$ yield will be modified by the dense medium is also not well quantified. The yield of the Drell-Yan and $b\bar{b}$ was parameterized as A/(1+m/m$_0$)$^n$ with n=4.69 and m$_0$ =2.7 as outlined in ref [4]. The line-shape of the ϒ(1S+2S+3S) was parameterized with three crystal ball functions representing the successive states. The mass values and their relative ratios were set to the values listed in the PDG. The width of the Gaussian piece and the size of the bremsstrahlung tail come from simulation. The like-sign subtracted invariant mass spectrum was then fit with a functional form including both of these pieces with two free parameters for the total yield of the background and of the ϒ(1S+2S+3S). This fit is shown in Figure 2. However, extracting the ϒ yield from this line-shape would bias the result, as the real line-shape of the ϒ within STAR depends on the levels of suppression which are not as yet quantified. The fit shown on Figure 2 was only used to determine the yield of the Drell-Yan and $b\bar{b}$ background. The ϒ yield used in the R$_{AA}$ calculation was determined by ϒ= N$_{+-}$ - N$_{--}$ - N$_{++}$ - ∫DY+ $b\bar{b}$. This process was repeated using only a single crystal ball function, representing the ϒ(1S) state as the most extreme possible line-shape and the difference in the two ϒ yields was included in the systematic uncertainty. The ϒ yield changed in these two circumstances because the yield of the Drell-Yan and $b\bar{b}$ from the fit did have some dependence on the ϒ line-shape.

The raw yield of ϒ→e+e- with |y| in the 0-60% centrality bin with |y$_ϒ$|<0.5 was determined to be 196.6 ± 35.8 (stat.). The efficiency times acceptance of this yield was determined by simulation. This was combined with the ϒ cross-section calculated in reference [4].

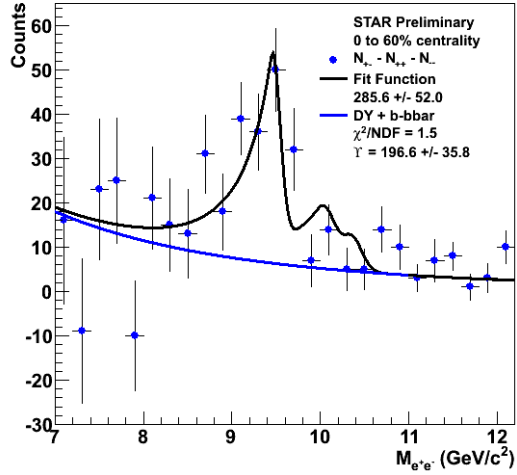

**Figure 2.** The blue points are the result from subtracting the like-sign yield from the unlike-sign yield shown in figure 1. The black curve is the combined line-shape that includes a parameterization of the ϒ(1S+2S+3S) + Drell-Yan + $b\bar{b}$ yield. The number of ϒs is determined by subtracting the integral of the blue curve from the histogram within the mass region of 8 to 11 GeV/c².

This results in an $R_{AA}$(0-60%)=0.56±0.11(stat)+0.02/-0.14(sys). This result does not contain the additional 14% systematic uncertainty and 33% statistical uncertainty from the p+p cross-section. We then further separated the yield into three bins in centrality: 0-10%, 10-30% and 30-60% as shown in Figure 3.

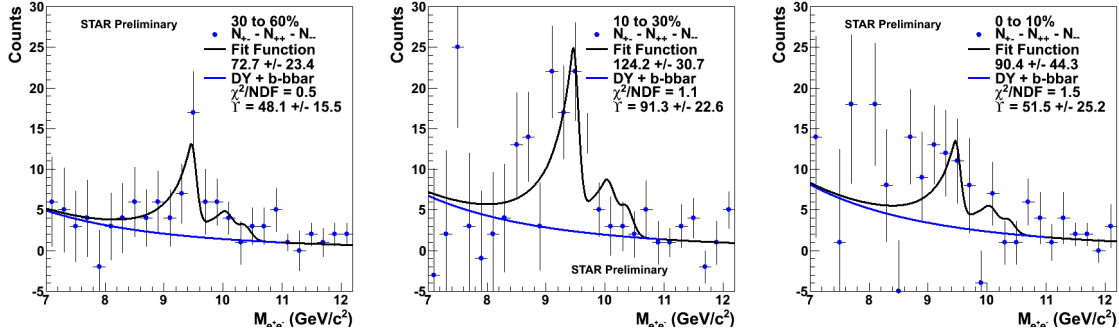

**Figure 3.** The graph on the left is the result from the 0-10% bin, the middle is from 10-30% and the right is 30-60%. The blue points are the result from subtracting the like-sign yield from the unlike-sign yield in each separate centrality bin. The black curve is the combined line-shape that includes a parameterization of the ϒ(1S+2S+3S) + Drell-Yan + $b\bar{b}$ yield. The number of ϒs in each graph is determined by subtracting the integral of the blue curve from the histogram within the mass region of 8 to 11 GeV/c².

The results from Figure 3, combined with efficiency corrections are shown in Figure 4. A clear trend versus centrality can be seen in this graph. Also shown are the high $p_T$ J/ψ results from STAR, the uncertainties due the ϒ p+p cross-section which would scale all three points.

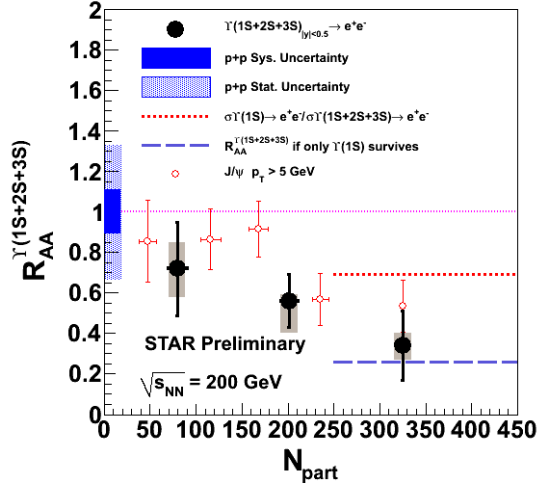

**Figure 4**. $R_{AA}$ for $\Upsilon(1S+3S+3S)_{|y|<0.5}$ versus centrality. The solid black points are the $\Upsilon$ results from Figure 3. The red open points are the high $p_T$ J/$\psi$ results from STAR, the details will be shown in these proceedings. The solid blue box is the systematic uncertainty from the p+p cross-section, resulting from the uncertainty in the luminosity and the trigger efficiency. The grey boxes around the three $\Upsilon$ points are the systematic uncertainties of those points resulting in the uncertainty in $N_{bin}$, the line-shape, and the Drell-Yan and $b\bar{b}$ yields. The red dotted line is the ratio of the total cross-section of $\Upsilon(1S)/\Upsilon(1S+2S+3S)$. The purple dashed line is the ratio of only the direct $\Upsilon(1S)$ cross-section over the total $\Upsilon(1S+2S+3S)$ cross-section.

## 3. Conclusions

The STAR experiment has measured the $R_{AA}$ for the 0-60% bin was calculated to be $0.56\pm0.21$(stat)$+0.08/-0.16$(sys) including systematic uncertainties from the p+p cross-section. Our data indicates $\Upsilon$ suppression in central collisions, where $R_{AA}$ is more than $3\sigma$ away from 1, including p+p uncertainties, in 0-10% centrality. A clear trend versus centrality can be observed.